\documentclass[aps,prl,twocolumn,groupedaddress]{revtex4}

\usepackage{graphicx}

\begin{document}

%\preprint{}

\title{Evolution of cooperation and communication skills as a consequence of environment fluctuations}


\author{A. Feigel}
\email[]{afeigel@fas.harvard.edu}
%\thanks{}
\altaffiliation{Dept. of Chemistry and Chemical Biology, Harvard University, 12 Oxford Street, Cambridge MA 02138 }
\affiliation{Center for Studies in Physics and Biology, The Rockefeller University, 1230 York Avenue, New York NY 10021, USA}


\date{\today}

\begin{abstract}
Dynamics of a social population is analyzed taking into account some physical constraints on individual behavior and decision making abilities. The model, based on Evolutionary Game Theory, predicts that a population has to pass through a series of different games, e.g as a consequence of environmental fluctuations, in order to develop social cooperation and communication skills. It differs from the general assumption that evolution of cooperation, the so called Cooperation Paradox, can be explained by a single set of rules for intra-population competitions. The developed methods, potentially, have a practical value for some learning optimization problems in multiagent, e.g. corporate, environment.
\end{abstract}

%\keywords{}

\maketitle

Behavior of social animals, especially human beings, is characterized by controversial equilibrium between selfishness\cite{daw76} and altruism, acting alone or joining an alliance, logical thinking and irrationality\cite{tk74}. Ambiguity of the observations complicates the separation of the field data into causes and consequences of the evolution of social behavior. Moreover, the relevant original phenomena can be suppressed in the modern populations. Therefore, it is beneficial to study the emergence of sociability presenting evolution as a game imposed on the individual members of a population and comparing the different possible evolutionary mechanisms using the Evolutionary Game Theory\cite{smi82}. 

The main question of the theory of social evolution is, probably, how the "selfish" competitions for the better individual share of the genome pool of the subsequent generation led to the development of willingness to contribute to the others on the personal expense? Any act committed in favor of the others may contradict to the Darwinian survival of the fittest. Indeed, Cooperation Paradox emerges trying to define the main properties of the game corresponding to the development of social cooperation: the games favoring single selfish winners, like Prisoners Dilemma or Hawk-Dove Games, correspond to our notion of the reality, although predict lower level of cooperation than it is believed to be present in nature.

There are three general possibilities to explain observations of behavior that seem to be irrational from the Darwinian evolution point of view. First, a help to the others can be justified by some unforeseen personal interest, e.g. because of group\cite{sob91,ric01,gin00} or kin (self-sacrifice in favor of the family)\cite{ham64,axe04} selections, as well as due to future award or punishment ensured by reciprocity\cite{tri71,ah81,ns05} or altruistic punishment\cite{ff03}. Second, one can assume that the optimum of individual behavior has not been evolved yet or its fluctuations are significant. Third, physical constraints on the possible evolutionary developments can cause cooperation-like behavior, e.g. enforcement of fair signalling by physical inability to deceive\cite{zah75}. The group selection is considered to be a weak phenomenon\cite{sob91}. It makes the physical constraints approach to be, probably, the only one capable of providing an explanation to cooperation between unrelated individuals with no common past or future, in a form of a general property of our world.

In this Letter a model of an evolving social population is constructed, based on, probably, the oldest social dilemma: acting alone or joining an alliance. The requirements for specific assumptions on the individual decision making mechanisms and structure of intra-population interactions network\cite{san05} are overcomed using symmetry considerations and some physical constraints, e.g. inability to divide resources like small amount of food or mating opportunities in mammals, together with limitations on the rational group decision making, discussed in social science and psychology\cite{str94,st85,col03}. The current state of a population is demonstrated to depend on the whole history of the evolutionary selection rules, rather than the most recent ones, providing an explanation for the observed significant differences in social abilities of otherwise similar groups. In addition a population based on mutual reciprocal exploitation, characterized by the correlated mutual responses of cooperate/exploit type, is predicted to be the most likely evolutionary development. The correlated responses of this type are impossible without development of information exchange between individual members of the population. The population dynamics and stability conditions are analyzed for a broad range of possible evolutionary games, including Prisoners Dilemma and Chicken (Snow-Drift, Hawk-Dove) Game. The results seem to correspond to the available qualitative experimental data\cite{roth88}. 

Quantitative description of the evolution of a population requires parametrization of the possible individual responses, inter-population interactions and individual decision making abilities. 
Fortunately, in biology the binary responses are common, for instance fight or retreat, cooperate or defect, etc. Therefore, we consider a population composed of individuals capable of generating only selfish $1$ and cooperative $2$ responses. Individuals compete with each other for some resources alone or creating the alliances. The price and the benefits of the competition are taken into account through the payoffs for choosing specific response against the response of the opponent (see Fig. 1A). The payoffs can vary with time favoring or suppressing the cooperation, e.g. due to the changes in value of the resources caused by environment fluctuations\cite{kl05}.

In the presented model, each individual possesses three parameters to evolve: sociability $\epsilon$ together with $\alpha$ and $\beta$, describing individual behavior in the course of a competition. The inter-population competitions are approximated by the pairwise interactions, with only one of the opponents acting either individually or as a member of an alliance, with individual probabilities $1-\epsilon$ and $\epsilon$ correspondingly (see Fig. 1B). It corresponds to the biological species competing for almost indivisible prizes, like small amounts of food or mating opportunities of mammals. In this case, each prize can be consumed only by the individual winner of individual/individual or individual/alliance contests. The parameter $\epsilon$ can be considered as a measure of the presence of the alliances: in a homogeneous population with $\epsilon=0$ all the competitions are of individual/individual type, otherwise if $\epsilon=1$ only individual/alliance contests occur.

An individual acting alone is characterized by the probabilities $\alpha$ and $\beta$ to be in cooperative response $2$ against the opponent in selfish $1$ and cooperative $2$ modes correspondingly (see Fig. 1C), receiving payoff according to the table of payoffs (see Fig. 1A). We assume that the members of a population have equal opportunities to consume specific amount of prizes during all possible interactions, rather than per single pairwise interaction. Therefore the table of payoffs describes the average total possible payoffs of an individual for its competitions in non-member of an alliance mode, although it requires competitions with prizes of different values to the competitors. The parameters $\alpha$ and $\beta$ are similar to the correlations with previous choice of the opponent\cite{em81,chr91,ns04}. However, $\alpha$ and $\beta$ include individual decision making abilities and any (rather than only memory based) detection of the intentions of the others. It corresponds to the biological species, especially human beings, recognizing the intentions of the others with no common past\cite{tcc+05}.

An individual acting as a member of an alliance generates random response according to the average statistics of its responses, and gets no payoff for the interaction. The random behavior is an extreme approximation of the possible constraints on the decision making abilities of an individual in a group\cite{str94,st85,col03}. It can be interpreted as some type of conformal behavior\cite{sky05}, taking into account that a random response, biased by the average behavior, is indistinguishable from the "average behavior" definition of social norms\cite{lev05}. The no payoff condition takes into account the reduction of the cost of the defeat and the benefit of the success for the members of an alliance, competing together for almost indivisible prizes. Surprisingly, these approximations make possible the optimization of the benefit of the whole population by the selfish competition between its members.

Evolution of a population consists of two processes: density redistribution between existing phenotypes $(\alpha_{i},\beta_{i},\epsilon_{i})$ and emergence of new mutations with initial minimal "seed" densities. The first process is described by Replicator Dynamics\cite{tay78} equations:
\begin{equation}
\frac{d}{{dt}}\rho_{i}=\rho_{i}\frac{F_{i}-\bar F}{{|\bar F|}},\;\bar F= \sum_{i=1}^N \rho_{i}F_{i}, \label{repdyn}%
\end{equation}
where $\rho_{i}$ is the density (sometimes called frequency) of the individual $i$ in the population. The individual fitness $F_{i}$ is the the averaged payoff $P_{i}$ of individual $i$ over all possible interactions. To define emergence of new mutations, the $(\alpha,\beta,\epsilon)$ space state density is assumed to be homogeneous and new mutations are assumed to be small modifications of the existing phenotypes\cite{em81,chr91,ns04}. 

The individual fitness $F_{i}$ (see eq. (\ref{repdyn})) depends on the all phenotypes $(\alpha_{j},\beta_{j},\epsilon_{j})$ present in the population. According to the definitions of $\alpha$ and $\beta$, the average payoff for an individual $i$ for the interaction with an opponent possessing probability $\gamma$ to be in the selfish state $1$ is (see Fig. 1A and 1C):
\begin{equation}
P_{i}(\gamma)=\gamma\alpha_{i}c-\gamma(1-\alpha_{i})+(1-\beta_{i})(1-\gamma)b, \label{deltaav}%
\end{equation}
where $b$ and $c$ are the payoffs (see Fig. 1A). 
Let us define $\gamma_{ij}$ to be the average $\gamma$ of individual $(\alpha_{i},\beta_{i})$ interacting with individual $(\alpha_{j},\beta_{j})$. In a population composed of identical individuals, self-consistent symmetry requires $\gamma_{ii}=(1-\alpha)\gamma_{ii}+(1-\beta)(1-\gamma_{ii})$, see Fig. 1C and 1D. Consequently:
\begin{equation}
\gamma_{ii}=(1-\beta)/(1+\alpha-\beta). \label{gammameanfield}%
\end{equation}
In the same fashion, $\gamma_{ij}=(1-\alpha_{i})\gamma_{ji}+(1-\beta_{i})(1-\gamma_{ji})$.
Taking into account the mirror equation for $\gamma_{ji}$ one gets:
\begin{equation}
\gamma_{ij}=((1-\beta_{i})-(1-\beta_{j})(\alpha_{i}-\beta_{i}))/(1-(\alpha_{i}-\beta_{i})(\alpha_{j}-\beta_{j})). \label{meanindgamma}%
\end{equation} 
The average $\gamma_{i}$, corresponding to the probability to find an individual $i$ in the state $1$, required to describe its response as a member of an alliance, is derived from self-consistent system of equations:
\begin{widetext}
%\begin{eqnarray}
\begin{equation}
\gamma_{i}=\sum_{j=1}^N \left[(1-\epsilon_{j})\rho_{j}((1-\alpha_{i})\gamma_{ji}+(1-\beta_{i})(1-\gamma_{ji}))+\epsilon_{j}\rho_{j}((1-\alpha_{i})\gamma_{j} +(1-\beta_{i})(1-\gamma_{j})\right], \label{realsys}%
%\end{eqnarray}
\end{equation}
\end{widetext}
where the averaging is performed over all interactions with all other individuals $j$, in their individual $1-\epsilon_{j}$ and member of an alliance $\epsilon_{j}$ modes. The fitness $F_{i}$ is obtained by averaging $P_{i}$ over all possible interactions, similar to eq. (\ref{realsys}): 
\begin{eqnarray}
F_{i}=\sum_{j=1}^N [(1-\epsilon_{j})\rho_{j}P_{i}(\gamma_{ji})+\epsilon_{j}\rho_{j}P_{i}(\gamma_{j})]. \label{realfit}%
\end{eqnarray}
Eqs. (\ref{repdyn}) and (\ref{realfit}) define the evolutionary dynamics of the proposed model. 

Surprisingly, the development of sociability $\epsilon$ is much slower process rather than the evolutionary dynamics of the individual behavior parameters $\alpha$ and $\beta$. For all possible games payoffs $b$ and $c$, there is no relative evolutionary advantage (any difference in fitness (\ref{realfit})) between two competing equipopulated subgroups, different only by the value of sociability: ($\alpha,\beta,\epsilon$) and ($\alpha,\beta,\epsilon+\Delta\epsilon$) (see supplementary materials). Consequently, there is no evolutionary preference for a change of the sociability of a confined population, unless its $\alpha$ and $\beta$ vary with time (see Fig. 2A). If $b$ and $c$ are constant, the total change $\Delta\epsilon_{total}\propto\Delta\epsilon_{mut}\sqrt{\Delta\alpha_{total}^{2}+\Delta\beta_{total}^{2}}\;$ ($\Delta\epsilon_{mut}$ is a single mutation step), remaining to be small, because the possible $\Delta\alpha_{total}$ and $\Delta\beta_{total}$ are limited by $\approx1$, until the convergence to one of the stable points. Therefore, the significant changes in $\epsilon$ can be caused only by alternating with time interaction payoffs $b$ and $c$. One example of the development of $\epsilon$ with time is shown in Fig. 2B, although the general problem of optimal $b(t)$ and $c(t)$ remains to be an open question.

The evolutionary stable values of $\alpha$ and $\beta$ can be treated as the functions of the game payoffs $b$ and $c$, together with sociability $\epsilon$, due to the slow dynamics of the latter (see Fig. 3). Only developed sociability $\epsilon=1$ ensures the stability of a population based on mutual reciprocal exploitation ($(\alpha=1,\beta=0)$, corresponding to the pairwise interactions of selfish vs. cooperative type $12$) for all range of the payoffs corresponding to the Chicken (Snow-Drift, Hawk-Dove) game (see Fig. 3A). Decrease of the value of $\epsilon$ reduces the number of games that can keep population out of homogeneous states (see Fig. 3). Evolution of a population with correlated mutual individual responses ($\alpha\neq\beta$, see Fig. 1D) is possible only with development of information exchange between the individuals\cite{aum76}. Consequently, it can be associated with development of communication skills, like language or writing. 

Evolution of biological species based entirely on Prisoners Dilemma, according to the model, results in a selfish population ($(\alpha=0,0\leq\beta<1)$, pairwise interactions $11$), see Figs. 3A and 3D. The individuals of this population, however, can possess an ability to recognize cooperation (state $2$) and cooperate in response ($\beta>0$, see Figs. 3D, 3A and 1D). The finite level of $(22)$ mutual responses, $\beta(1-\gamma)\neq0$,  must be observed in the population with $\gamma<1$ and $\beta>0$, playing Prisoners Dilemma for a period of time that is too short to change the individual parameters. It corresponds qualitatively to the experiments where different population demonstrate different level of cooperation\cite{gin04} and degradation of cooperation with time is observed\cite{roth88}.

The model predicts the optimal games for the fastest transition from a selfish or cooperative populations to an exploitative one (see Fig. 4), in case of developed sociability $\epsilon=1$. The time required to develop specific population depends on the properties of the population itself and the history of the game parameters $b(t)$ and $c(t)$. It is a consequence of the topological constraints on the dynamics in $(\alpha,\beta,\epsilon)$ space, defining the mutations that can take over the population. For instance, much longer time required to develop synchronous population ($(\alpha=0,\beta=1)$,pairwise interactions $11,22$), rather than exploitative one ($(\alpha=1$,$\beta=0$, pairwise interactions $12$), due to the small amount of the allowed mutations near the axis $\alpha=0$ and $\beta=1$ in this case (see Fig. 3). The synchronization of the responses, like in synchronous population, is common between the cells in the multicellular organisms. Therefore this prediction is intriguing, due to disproportional large time taken for development of multicellularity, relative to the other major evolutionary transitions\cite{ss95}. The existence of the optimal game corresponds to our intuition that neither too harsh nor too soft conditions are optimal for a learning process.

To conclude, the Cooperation Paradox was addressed by taking into account the advantages and disadvantages of being a member of an alliance, especially the possible constraints on the individual decision making abilities. It was demonstrated that a population has to pass through a series of different conditions, favoring and suppressing selfishness, in order to develop a robust sociability based on contribution to the others at the personal expense. In case of developed sociability, an exploitative population, characterized by the correlated mutual responses of cooperate/exploit (rather than homogeneous cooperation or selfishness) type, was shown to be stable for a whole range of the Chicken (Hawk-Dove, Snow-Drift) Game.
The presented method, being free from the specific assumptions on the individual decision making mechanisms, provides a general framework for analysis of different hypothesis of evolution of social behavior. The future developments of the model can include modification of the inter-population interactions rules and investigation of the possibility of the individuals affect the abilities of the others, e.g. evolution of the ability to deceive.
%\subsection{}
%\subsubsection{}

\section{Acknowledgments}
The author is grateful to A. Libchaber, M. Aizenman, G. Falkovich, M. Krieger, S. Leibler, A. Morozov, M. Narovlyansky, A. Roth, B. Skyrms and A. Zilman for useful discussions.
\bibliography{corevol1}
\newpage
\begin{figure}
\includegraphics{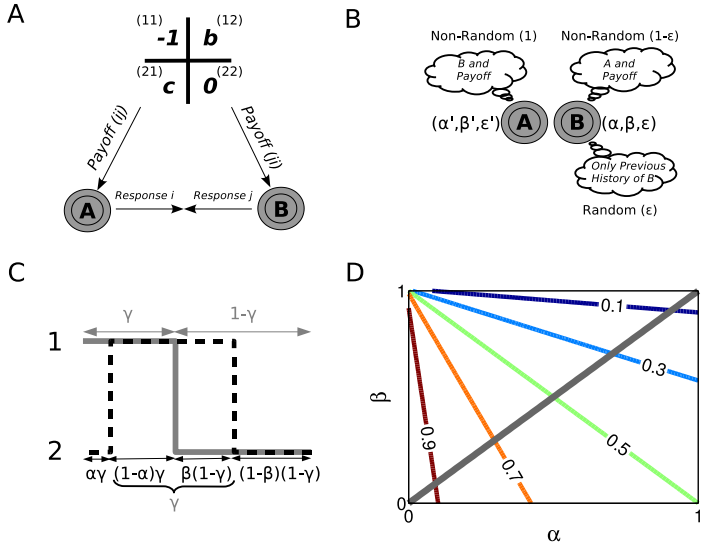}%
\caption{{\bf The evolution of a population is defined by the interactions on the individual level}. ({\bf A}), Table of payoffs $W_{ij}$ for the pairwise interactions. The table is reduced to two parameters form by subtraction of $W_{22}$ and subsequent normalization by $|W_{11}-W_{22}|$. Two, rather than four, parameters significantly simplify the presentation. This normalization does not affect the stable points of the population dynamics (see eqs. (\ref{repdyn}) and (\ref{deltaav})). ({\bf B}), The asymmetric probability to generate random response during an interaction $\epsilon$, interpreted as acting alone or out of an alliance. ({\bf C}), Correlations $\alpha$ and $\beta$ describe the individual (dashed line) ability to recognize the intentions of the opponent (solid line) and make an appropriate decision. The parameter $\gamma$ corresponds to the probability of finding an individual in the selfish response $1$. ({\bf D}), The $\gamma$ of a homogeneous population composed of individuals $(\alpha,\beta,\epsilon)$. The points $(0,1)$ and $(1,0)$ describe the correlated populations with set of mutual responses $(11,22)$ and $(12)$ correspondingly. The condition $\alpha=\beta$ corresponds to the populations composed of individuals randomly choosing selfish or cooperative responses, disregard of the state of the opponent.}
\end{figure}
%\newpage
\begin{figure}
\includegraphics{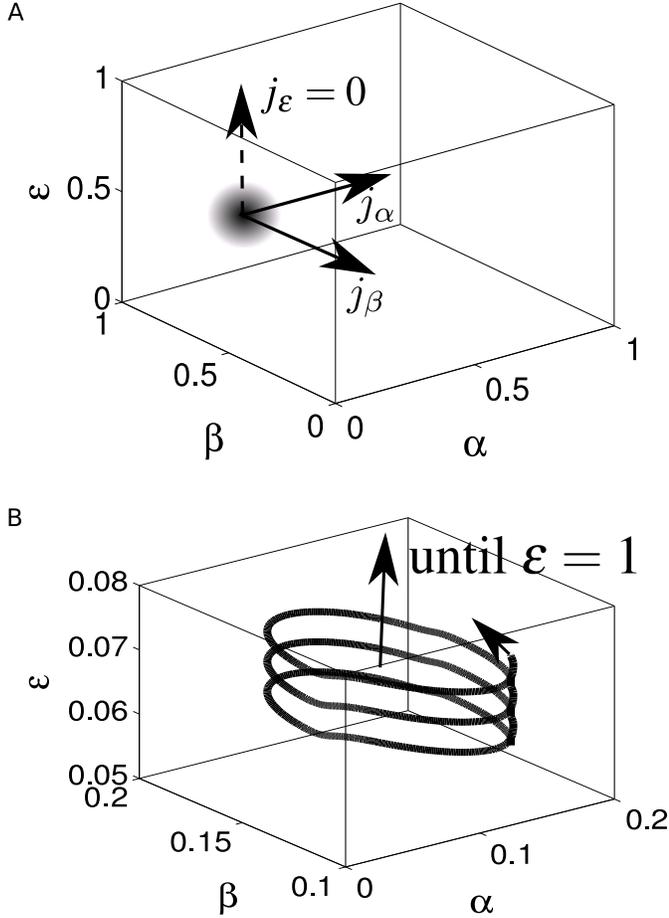}%
\caption{{\bf Dynamics of a confined population in $(\alpha,\beta,\epsilon)$ space is governed by the time dependent game payoffs $b(t)$ and $c(t)$.} ({\bf A}), A population is driven by the game payoffs, making specific mutations to do better than the others. The population remains confined, while new mutations appear less frequent rather than the old take over the populations according to eq. (\ref{repdyn}). Surprisingly, a confined population in $(\alpha,\beta,\epsilon)$ space experiences first order drag, defined by the game weights $b$ and $c$, only in $(\alpha,\beta)$ plane. The development of the social behavior $\epsilon$ is the second order process, requiring $\Delta\alpha,\Delta\beta\neq0$.  ({\bf B}), The development of $\epsilon$ requires continuous motion away of the points where both $\alpha$ and $\beta$ are constant. Here is an example of $\epsilon(t)$ with $\Delta\epsilon_{mut}=0.05$, one mutation step in $time=10$, $b=1+2cos(2\pi 10^{-3}N_{steps})$, $c=2sin(2\pi 10^{-3}N_{steps})$ and $\alpha,\beta$ confined to the circle with the center at $(0.11,0.15)$ and the radius $0.05$. The $(\partial\alpha/\partial t,\partial\beta/\partial t)$ was derived numerically and used to estimate $\partial\epsilon/\partial t$. The change of the game payoffs can be a consequence of both individual evolution or environment changes. Evolution of life on Earth, oscillating from harsh to more soft conditions, justifies the requirement to consider time dependent $b(t)$ and $c(t)$.}
\end{figure}
%\newpage
\begin{figure}
\includegraphics{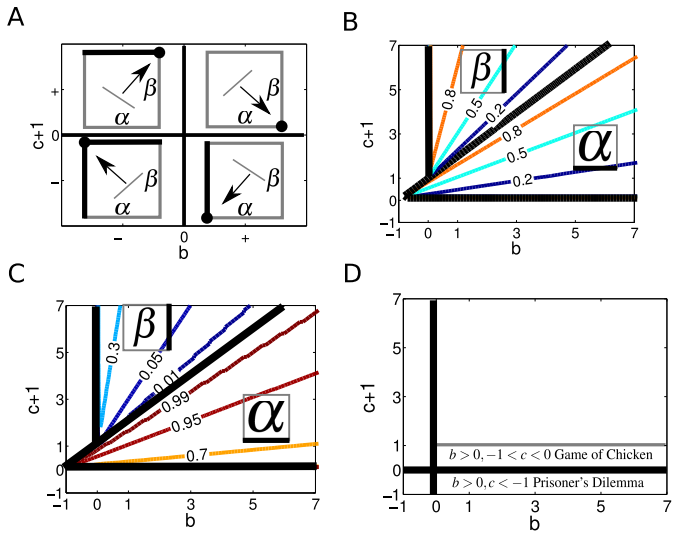}%
\caption{{\bf The stable points in $(\alpha,\beta,\epsilon)$ space.}
({\bf A}), In the case $\epsilon=1$ for all individuals in the population, all of them experience the rest of society as an individual in the alliance mode, described by the probability $\gamma_{ii}$ (see eq. (\ref{gammameanfield})) to express response $1$. Mutant $j$ can invade the population of $i$ only in case $P_{j}(\gamma_{ii})>P_{i}(\gamma_{ii})$. Using eq. (\ref{deltaav}) one can write $\Delta\beta(1-\gamma_{ii})b<\Delta\alpha\gamma_{ii}(c+1)$, where $\Delta\alpha=\alpha_{j}-\alpha_{i}$ and  $\Delta\beta=\beta_{j}-\beta_{i}$. Near the axis $\alpha=0$ and $\beta=1$, for specific $b$ and $c$, the amount of the evolutionary favorable mutations is converging to $0$, see $\gamma_{ii}$ in Fig 1D, making the dynamics to be very slow. Numerical simulation demonstrate the same behavior for all range of $0<\epsilon<1$. It makes the exploitative population $(\alpha=1,\beta=0)$ to be the only one that can be reached in a reproducible way. The exploitative population is more beneficial on average to its members rather than cooperative ($(0<\alpha\leq 1,\beta=1)$,$22$) one, if $(b+c)/2>0$. Otherwise, if $(b+c)/2<0$, the exploitative population is still evolutionary stable, although the cooperative society is more beneficial. ({\bf B}), In the case $\epsilon=0$, the system either converges to a specific points on the axis $\beta=0$ and $\alpha=1$, or, otherwise, remains at random point on one of the edges $\alpha=0$ or $\beta=1$. The presented data was derived using the condition for the stable points on the boundaries and direct simulation of populations composed of up to $10$ different phenotypes $(\alpha,\beta,\epsilon)$. ({\bf C}), The same for the case $\epsilon=0.9$. ({\bf D}), Comparison with the model games demonstrate that the population playing the Chicken Game (sometimes called as Hawk-Dove or Snow-Drift, $W_{12}>W_{22}>W_{11}>W_{21}$), rather than Prisoner's Dilemma ($W_{12}>W_{22}>W_{21}>W_{11}$), is able to stabilize the exploitative population $(1,0)$.}
\end{figure}
%\newpage
\begin{figure}
\includegraphics{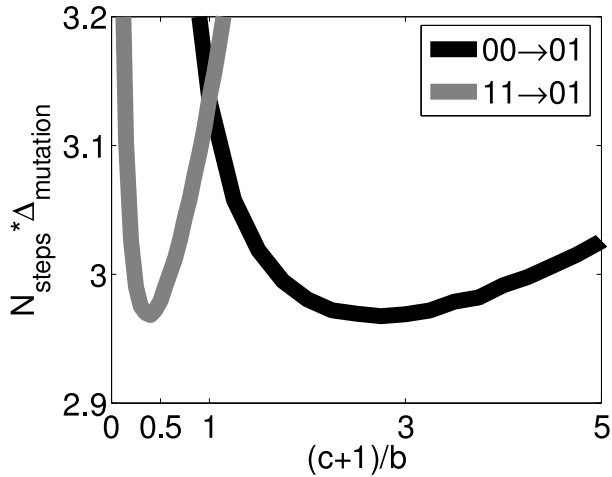}%
\caption{{\bf Development of mutual correlation in the population with $\epsilon=1$ as a function of the game payoffs $b$ and $c$.}
The time required for the transition from a population with random mutual responses to the correlated one, was analyzed by numerical simulation. Such transition can occur during the development of a common language. Fast mutation spread over the entire population was assumed, corresponding to the cultural evolution. In the case $\epsilon=1$, following condition $\Delta\beta(1-\gamma_{ii})b<\Delta\alpha\gamma_{ii}(c+1)$ (see Fig. 3A), the $(c+1)/b$ is the only relevant parameter and two transition, $(0,0)\rightarrow(1,0)$ and $(1,1)\rightarrow(1,0)$, are identical under $(c+1)/b\rightarrow b/(c+1)$ transformation. The group selection can be introduced by requirement for growing average fitness of the population, predicting the optimal games at $(c=1.5\pm0.01,b=0.9\pm0.03)$ and $b\rightarrow\infty$, for the transitions $(0,0)\rightarrow(1,0)$ and $(1,1)\rightarrow(1,0)$ correspondingly. The presented results are derived using single step diffusion simulation. The minima's search functions of GSL were used to find optimal values of $b$, $c$ and $(c+1)/b$.}
\end{figure}

\end{document}